\documentclass[12pt]{article}

\usepackage{english,a4}

\usepackage{epsfig}
\usepackage{dcolumn}
\usepackage{amsmath}
\usepackage{cite}
\usepackage{changebar}

\makeatletter \@addtoreset{equation}{section}
\makeatother 
\renewcommand{\thefootnote}{\alph{footnote}}

\newcommand{\scr}[1] {\mbox{\scriptsize #1}}

\newcommand{\DT} {\ensuremath{\Delta(T)\times T^{4}}}

\newcommand{\tc} {\ensuremath{T_{\scr{c}}}}

\begin{document}
\thispagestyle{empty}
%

\begin{center}
\vspace*{1.0cm}
\renewcommand{\thefootnote}{\fnsymbol{footnote}}
{\LARGE\bf  THE GLUON CONDENSATE IN QCD~\\ AT FINITE TEMPERATURE
\footnote{Presented at the XXXVII Cracow School of Theoretical Physics,
 Zakopane, Poland, May 30-June 10,1997} \\}

\vspace*{1.0cm}
{\large David E. Miller$^{1,2,3}$
\\}
\vspace*{1.0cm}
${}^1$ {Institute of Theoretical Physics, Plac Maksa Borna 9,
University of Wroclaw,\\
PL-50-204 Wroclaw, Poland}\\
${}^2$ {Fakult\"at f\"ur Physik, Universit\"at Bielefeld, Postfach 100131,\\ 
D-33501 Bielefeld, Germany}\\
${}^3$ {Department of Physics, Pennsylvania State University,
Hazleton Campus,\\
Hazleton, Pennsylvania 18201, USA (permanent address)}\\
\vspace*{2cm}
{\large \bf Abstract \\}
~\\
We begin with the discussion of the relationship between the trace 
of the energy momentum tensor and the gluon condensate at finite
temperatures. Using the recent numerical data from the 
simulations of lattice gauge theory for quantum chromodynamics(QCD)
we present the computational evaluations for the gluon condensate.
A short discussion of the properties of deconfinement and the
implications on the high temperature limit are included.
We also  mention the case of the massive quarks where some
of the properties of the condensate appear to change. We put together
these results with some ideas related to the dilatation current.
We draw the conclusion that the nature of the strong interactions implies
that the thermodynamics of quarks and gluons never approach even at
very high temperatures that of an ideal ultrarelativistic gas.
\end{center}
appeared in {{\it Acta Physica Polonica} {\bf B28}, 2937 (1997), corrected
 version  }
\newpage


\section{\bf Introduction}
\begin{center}
     {\it Out of the present findings based on previous knowledge
     new science is revealed.}
\end{center}
~\\
\noindent
This statement particularly pertains to the recent experimental results on the
Bose-Einstein condensation of atomic alkalai vapors~\cite{Andn}.
Although this talk will not be directly related to these results,
it draws much inspiration from them.
~\\
\indent 
     In this presentation we discuss the consequences for the gluon
condensate at finite temperature of the recent high precision lattice
results~\cite{Eng4,Boyd} for the equation of state in lattice gauge theory.
We compare these results with various other calculations of the
expected high temperature behavior. The essential relationship is the 
trace anomaly that is due to the scale variance of
quantum chromodynamics (QCD). It relates the trace of the
energy momentum tensor to the square of the gluon field strenths
through the renormalization group beta function. Here we shall
expand upon the approach investigated in~\cite{Mill}, for which the
consequences of the new finite temperature lattice data for $SU(N_{c})$
gauge theory for the gluon condensate~\cite{BoMi} have been presented.


\section{\bf The Trace Anomaly at finite Temperature}

\indent
     The idea of the relationship between the trace of the energy momentum
tensor and the gluon condensate has been studied for finite temperature 
by Leutwyler~\cite{Leut} in relation to the problems of deconfinement 
and chiral symmetry. He starts with a detailed discussion of the 
trace anomaly based on the interaction between Goldstone bosons 
in chiral perturbation theory. Central to his discussion is the 
role of the energy momentum tensor, whose trace is directly related
to the gluon field strength. It is important to note that the
energy momentum tensor $T^{\mu\nu}(T)$ can be separated into the zero 
temperature or confined part, $T^{\mu\nu}_{0}$, and the finite temperature
contribution $\theta^{\mu\nu}(T)$ as follows:
\begin{equation}
  \label{eq:emtensor}
  T^{\mu\nu}(T) = T^{\mu\nu}_{0} + \theta^{\mu\nu}(T) .
\end{equation}
\noindent
The zero temperature part, $T^{\mu\nu}_{0}$, has the standard problems with
infinities of any ground state. It has been discussed by 
Shifman, Vainshtein and Zakharov \cite{SVZ1} in relation to the 
nonperturbative effects in QCD and the operator product expansion. 
In what follows we shall just use a bag type of model~\cite{Chod}
as a means of stepping around these difficulties since we are only interested 
here in the thermal properties. The finite temperature part, 
which is zero at $T=0$, is free of such problems. We shall see in the next 
section how the diagonal elements of $\theta^{\mu\nu}(T)$ are calculated in 
a straightforward way on the lattice. The trace  $\theta^{\mu}_{\mu}(T)$ is
connected to  the thermodynamical contribution to the energy density 
$\epsilon(T)$ and pressure $p(T)$ for relativistic fields~\cite{LaL2} 
and relativistic hydrodynamics~\cite{LaL6}
\begin{equation}
  \theta^{\mu}_{\mu}(T) = \epsilon(T) - 3p(T) .
  \label{eq:eps-ideal}
\end{equation}
\noindent
The gluon field strength tensor is denoted by
$G^{\mu\nu}_a$, where $a$ is the color index for $SU(N)$.
The basic equation for the relationship between the gluon condensate
and the trace of the energy momentum tensor at finite temperature was
written down by Leutwyler \cite{Leut} using the trace anomaly in the form
\begin{equation}
\langle G^2 \rangle_{T} = \langle G^2 \rangle_0~-~
\langle {\theta}^{\mu}_{\mu} \rangle_{T}, 
\label{eq:condef}
\end{equation}
\noindent
where the gluon field strength squared summed over the colors is
\begin{equation}
G^{2}~=~{{-\beta(g)}\over{2g^3}} G^{{\mu}{\nu}}_{a}G_{{\mu}{\nu}}^{a}, 
\end{equation}
\noindent
for which the brackets with the subscript $T$ mean thermal average.
The renormalization group beta function $\beta(g)$ in terms of the
coupling may be written as
\begin{equation}
\beta(g)~=~\mu{dg \over{d\mu}}
         =~-{1 \over{48\pi^{2}}}(11N_{c}~-~2N_{f})g^{3}~+~O(g^{5}).
\label{eq:betafun}
\end{equation}
\noindent
He has calculated \cite{Leut} for two massless quarks using the low temperature
chiral expansion the trace of the energy momentum tensor at finite temperature
in the following form:
\begin{equation}
\langle \theta^{\mu}_{\mu} \rangle_{T}~=~-{\pi^{2} \over {270}}
{T^{8} \over {F^{4}_{\pi}}}{\left\{ln{{\Lambda_{p}}\over{T}}~-~{1\over4}\right\}}
~+~O(T^{10}),
\end{equation}
\noindent
where the logarithmic scale factor $\Lambda_{p}$ is about $0.275 GeV$ and
the pion decay constant $F_{\pi}$ has the value of $0.093 GeV$. The value
of the gluon condensate for the vacuum $\langle G^{2} \rangle_{0}$ was
taken to be about $ 2 GeV/fm^{3}$, which is consistent with the previously
calculated values \cite{SVZ1}. The results sketched by Leutwyler at QUARK
MATTER'96 in Heidelberg \cite{Leut} show a long flat region for
$\langle G^{2} \rangle_{T}$ as a function of the temperature until it
arrives at values of at least $0.1 GeV$ where it begins to show a falloff
from the vacuum value proportional to the power $T^{8}$.



\section{\bf Lattice Data for the Equation of State}

The lattice calculation at finite temperature proceeds 
(talk by Krzysztof Redlich \cite{Redl}) in the following way. From the action
expectation value at zero temperature, $P_{0}$, as well as the spatial and
temporal action expectation values at finite temperature, $P_{\sigma}$ and
$P_{\tau}$ respectively and $N_{\tau}$ the number of temporal steps, the 
dimensionless interaction measure $\Delta(T)$~\cite{Eng1,Eng2,Eng3} is given 
by
\begin{equation}
  \Delta(T) = -6N{N^{4}_{\tau}}a\frac{dg^{-2}}{da}\left[
                                2P_{0} - (P_{\sigma}+P_{\tau})
                                \right].
\end{equation}
\noindent
The crucial part of these recent calculations is the use of the full
lattice beta function, $\beta_{\scr{fn}}=adg^{-2}/da$
in obtaining the lattice spacing $a$, or scale of the simulation,
from the coupling $g^{2}$. Without this accurate information
on the temperature scale in lattice units it would not be possible
to make any claims about the behavior of the gluon condensate.
The dimensionless interaction measure is equal to
the thermal ensemble expectation value of $(\epsilon - 3p)/T^4$. Thus by the
equation~\eqref{eq:eps-ideal} above is equal to the expectation value
of the trace of the temperature dependent part of the energy momentum tensor,
~\cite{Mill} which may be written (after suppressing the brakets)as follows:
\begin{equation}
  \theta^{\mu}_{\mu}(T)~=~\DT.
\label{eq:trace}
\end{equation}
\noindent
There are no other contributions to the trace for QCD on the lattice. 
The heat conductivity is zero. Since there are no non-zero conserved quantum
numbers and, as well, no velocity gradient in the lattice computations, 
hence no contributions from the viscosity terms appear.
For a scale invariant system, such as a gas of free massless particles,
the trace of the energy momentum tensor, equation~\eqref{eq:trace}, is zero.
A system that is scale variant, perhaps from a particle mass, 
has a finite trace, with the value of the 
trace measuring the magnitude of scale breaking.
At zero temperature it 
has been well understood from Shifman et al.~\cite{SVZ1}
how in the QCD vacuum the trace of the energy momentum tensor
relates to the gluon field strength squared, $G^{2}_0$.
Since the scale breaking in QCD occurs explicitly at all orders in a loop
expansion, the thermal average of the trace of the energy momentum tensor
should not go to zero above the deconfinement transition.
So a finite temperature gluon condensate $G^{2}(T)$
related to the degree of scale breaking at all temperatures, can be defined
to be equal to the trace. We have used~\cite{BoMi} the lattice simulations
~\cite{Eng4,Boyd} in order to get the temperature dependent part 
of the trace and, thereby, the value of the condensate at finite temperature.
In what follows we will use the assumptions of a bag type of model~\cite{Chod}
which includes a bag pressure $-B$ and an energy density of the confined 
state of the same magnitude $B$. These assumptions mean that this
contribution to the trace of the energy momentum tensor becomes simply
${T_0}^{\mu}_{\mu}~=~4B$. Using this form we may write the equation of
state for the {\it total} energy density $\epsilon(T)$ and pressure $p(T)$
as follows:
\begin{equation}
\epsilon(T)~-~3p(T)~=~T^{\mu}_{\mu}(T)  
\label{eq:state}
\end{equation}
The trace of the energy momentum tensor as a function of the temperature
is shown in Figure 1. We notice that for $T~<~\tc$ it remains constant
at $4B$. However, above $\tc$  in both cases there is a rapid rise in
$T^{\mu}_{\mu}(T)$. 
Accordingly, the vacuum gluon condensate $G^{2}_{0}$ becomes just $4B$. Here
we have assumed the value $4B~=~0.012 GeV^4$  for both cases~\cite{SVZ1}.
By taking the published data~\cite{Eng4,Boyd} for $\Delta(T)$, and using
equations~\eqref{eq:condef} and~\eqref{eq:trace} we obtained the gluon
condensate $G^{2}(T)$ as shown in Figure 1.
\begin{figure}[b]
   \begin{minipage}{75mm}
  \begin{center}
    \leavevmode
    \epsfig{file=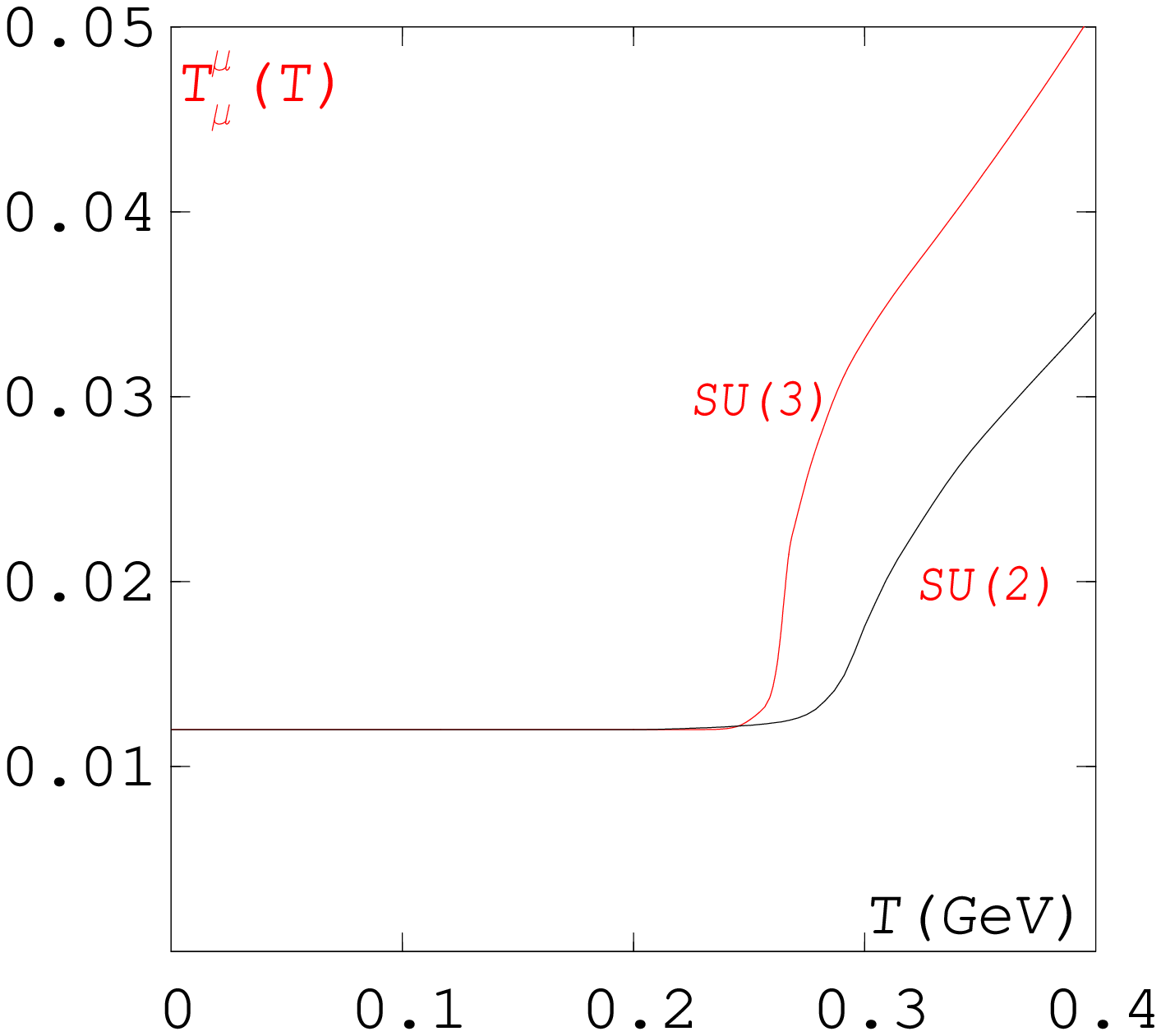
        ,height=85mm
        ,bbllx=85,bblly=200,bburx=550,bbury=650}
   \end{center}
   \end{minipage}
\hfill   
\begin{minipage}{75mm}
  \begin{center}
    \leavevmode
      \epsfig{file=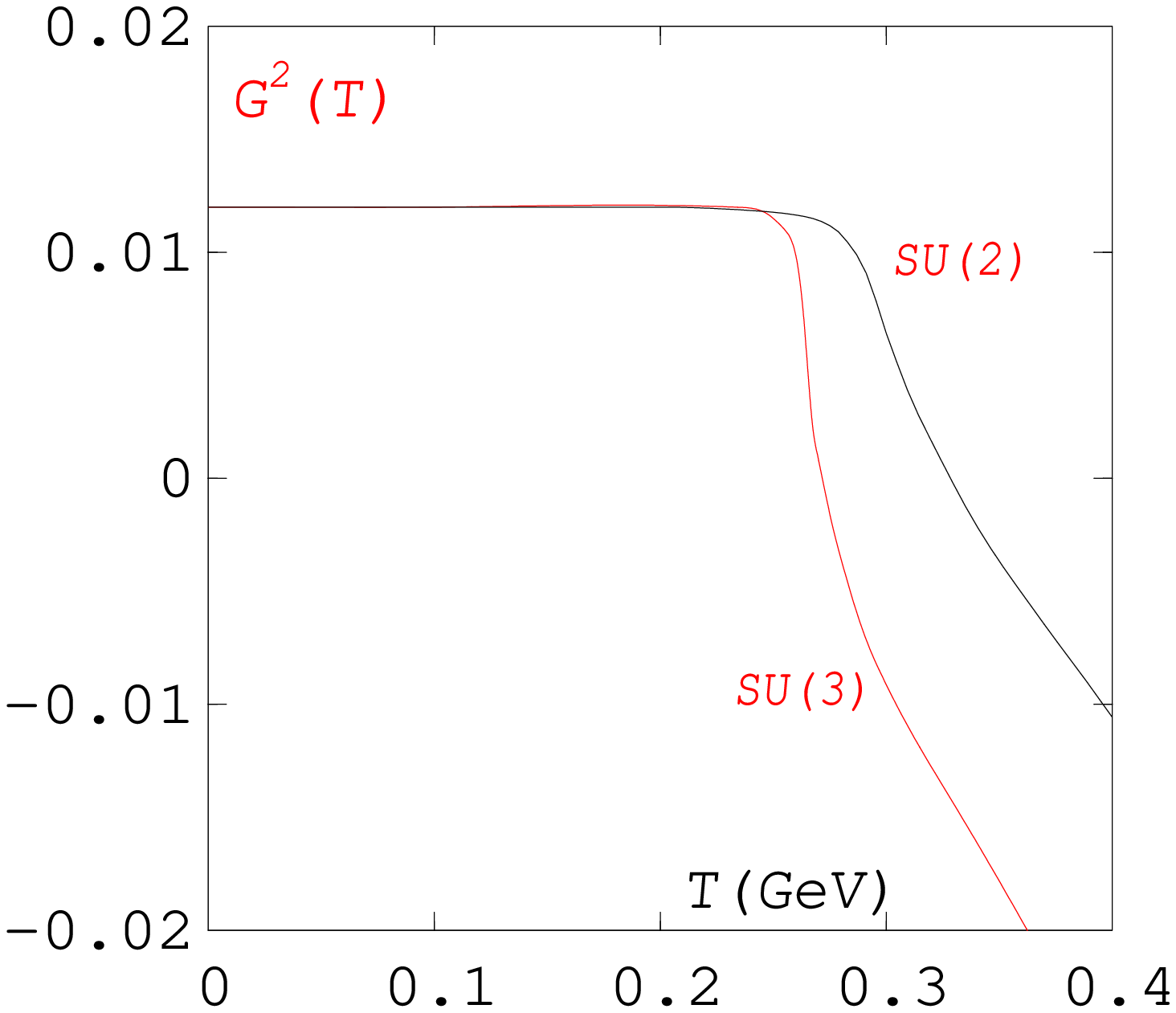,bbllx=85,bblly=200,bburx=550,bbury=650,
          height=85mm}
     \end{center}
   \end{minipage}
   \caption{      \label{fig:cond3}
   The left plots show $T^{\mu}_{\mu}(T)$ for the lattice gauge theories
   $SU(2)$ and $SU(3)$ as indicated. The right plots show the
   corresponding gluon condensates. The values for the critical temperature
   $\tc = 0.290, 0.264$~GeV for $SU(2)$ and $SU(3)$, respectively.
          }
\end{figure}



\section{High Temperature Behavior of the Equation of State}

In this section we present a discussion of the properties of the equation of
state at temperatures above $2T_c$ up to around $5T_c$.
If one were to consider a simple bag type of model with the bag constant B
independent of the temperature, one finds immediately the simple equality
\begin{equation}
\epsilon~-~3p~=~4B.
\end{equation}
This form is in clear conflict with what we found above as shown in
Figure 1(a), for which there was a steady rise in in its value.
Some of the early work in this temperature range was done using perturbative
estimates by K\"allman~\cite{Kall} and Gorenstein and Mogilevsky \cite{GoMo}.
These authors found essentially a linear rise in the equation of state a 
function of the temperature. Also Montvay and Pietarinen \cite{MoPi}
looked at the asymptotic properties of the gluon gas.
The asymptotic behavior of the
form of $\Delta(T)$ assumed by K\"allman \cite{Kall} and Gorenstein and
Mogilevsky \cite{GoMo} are qualitatively similar to our plots in as far as
the data extends. However, the latter \cite{GoMo} represent their data in
terms of energy density and the pressure separately showing how the two curves
converge to eachother at small values of $1/T^3$.
All these results are quite similar to the ideal relativistic massive gas.
The equation of state is then
\begin{equation}
\epsilon~-~3p~=~{1\over{2 \pi^2}}m^{3}T K_{1}{\left({m\over{T}}\right)}.
\end{equation}
We can readily see form the asymptotic properties of the modified Bessel
function $K_{\nu}(x)$ in this equation that the equation of state is always
positive and growing as a function of the temperature. In the high
temperature limit the equation of state relates to $(mT)^{2}$.

\begin{figure}[b]
   \begin{minipage}{75mm}
  \begin{center}
    \leavevmode
      \epsfig{file=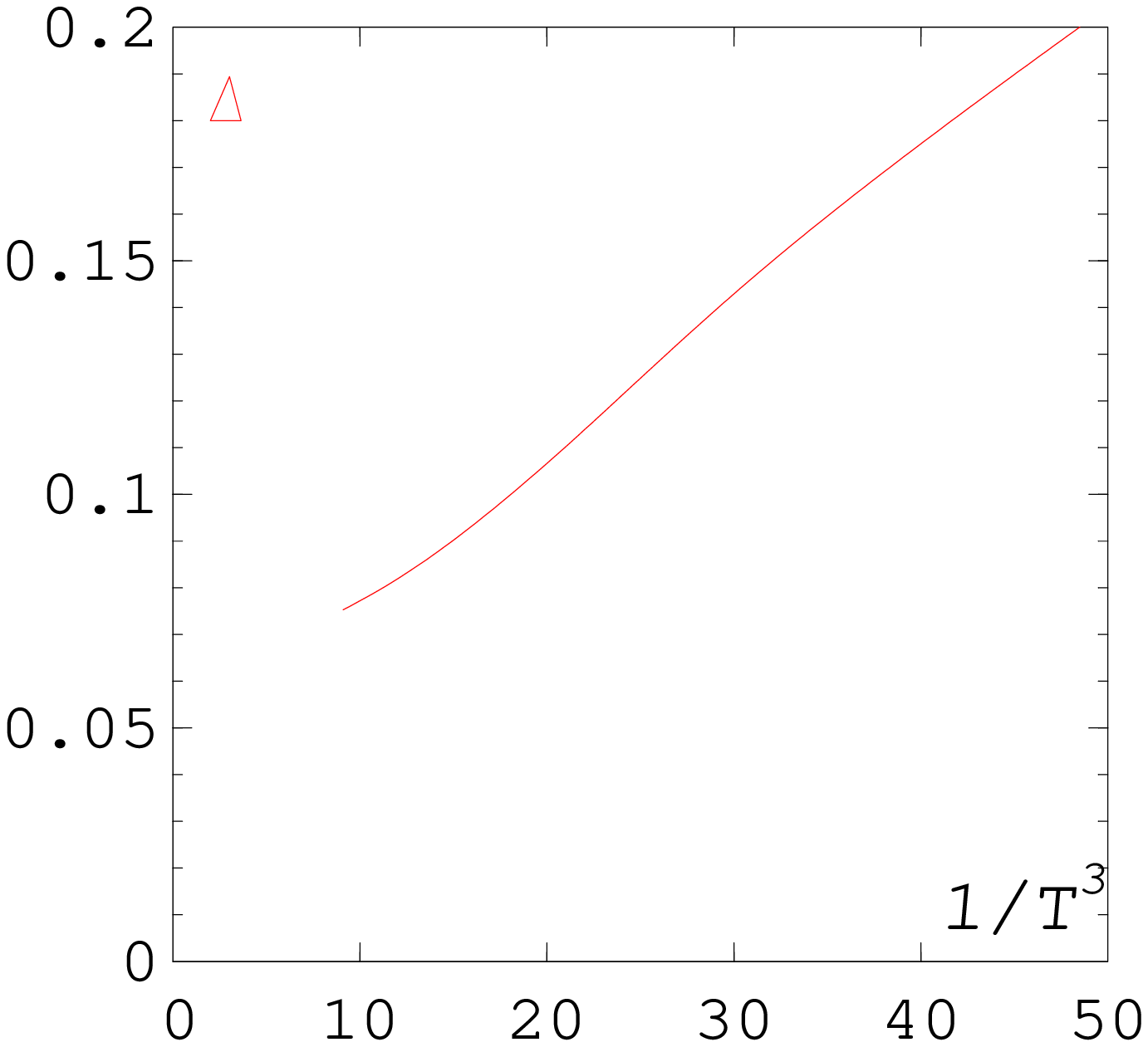,bbllx=85,bblly=200,bburx=550,bbury=650,
          height=85mm}
   \end{center}
   \end{minipage}
\hfill   
\begin{minipage}{75mm}
  \begin{center}
    \leavevmode
      \epsfig{file=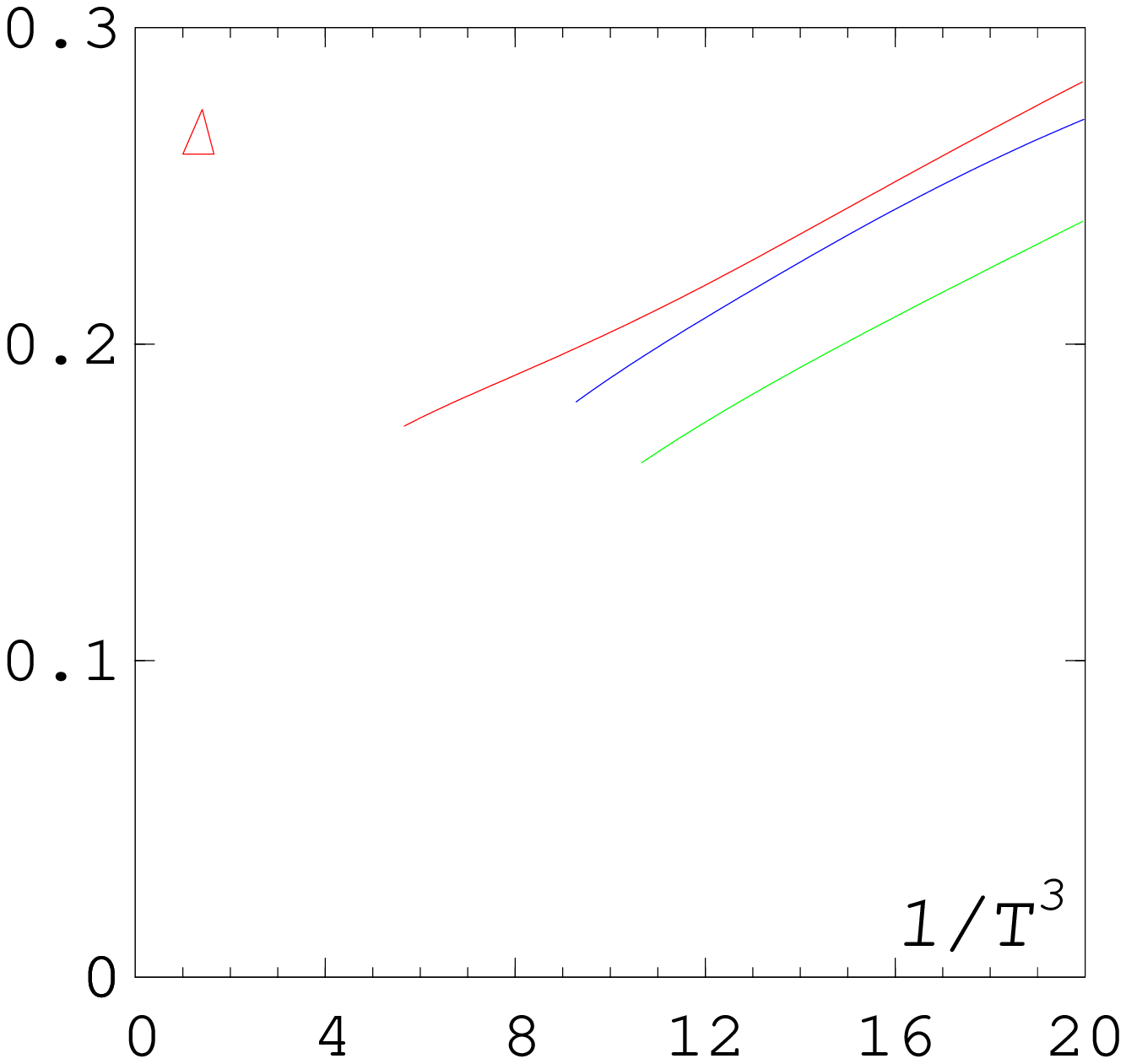,bbllx=85,bblly=200,bburx=550,bbury=650,
          height=85mm}
     \end{center}
   \end{minipage}
   \caption{      \label{fig:del_t3}
Figure~(a) shows $\Delta(T)$ as a function of $1/T^3$ for $SU(2)$ lattice
gauge theory\cite{Eng4},where the scale of the abscissa is ${\tc}^{-3}\times
10^{-3}$;
Figure~(b) shows $\Delta(T)$ for $SU(3)$ with the three lattice sizes \cite{Boyd} 
$16^3\times4$ (top), $32^3\times6$ (middle), $32^3\times8$ (bottom). The
scale of the abscissa is the same as for $SU(2)$.   }
\end{figure}
~\\
\indent
We plot here for the sake of comparison with these authors
the high temperature behavior of $\Delta(T)$ at temperatures in the above
range.  The following two plots show the high temperature behavior of $SU(2)$
and $SU(3)$ explicitly as a function of $1/T^3$.
We see from these plots that the general appearance is much the same in all the
different cases. Roughly speaking we could interpret $\Delta(T)$ as a linearly
increasing function in the variable $1/T^3$.
In Figure~\ref{fig:del_t3}(b) we have provided three lattice sizes for the
sake of comparison.  We notice that there exists a considerable gap with no
indication of $\Delta(T)$ approaching zero at small $1/T^3$.
Although we are not able to exactly determine the asymptotic form of
the function at high temperature, we are able to conclude about certain
behavior from given models in the range of temperature
between $2T_c$ and $5T_c$.
~\\
\indent
Finally we briefly mention the results of finite temperature  perturbation
theory~\cite{ArZh}. From the expression for the free energy one can derive
the interaction measure in the form
\begin{equation}
\Delta~=~{11\over{6}}C^{2}_{A}d_{A}{\alpha}^{2}{\left({1\over{36}}~+~
         {1\over{3}}\left({{C_{A}\alpha}\over{3\pi}}\right)^{1/2}~+~
         O(\alpha)\right)},
\label{eq:pert}
\end{equation}
where $\alpha$ is related to the coupling by $g^{2}/{4\pi}$. The constants
$C_{A}$ and $d_{A}$ relate to the number of colors as $N_{c}$ and $N^{2}_{c}
~-~1$ respectively. The significance of the factor $11\over{6}$ was pointed
out to us by Bo Andersson \cite{BoAn}as the value in the rapidity
of the gluon splitting. Thus we see from~\eqref{eq:pert} that the 
temperature dependence of the equation of state is very strong 
in comparison with our plots.



\section{Gluon Condensate with Quarks}

In this section we would like to discuss the changes due to the presence of 
dynamical quarks with a finite mass.  There have been recent computations of 
the thermodynamical quantities in full QCD with two flavors of staggered
quarks~\cite{BKT94,MILCeos6}, and with four flavors~\cite{edwinqmfklat96}.
These calculations are not yet as accurate as those in pure gauge theory for
two reasons. The first is the prohibitive cost of obtaining statistics similar
to those obtained for pure QCD. So the error on the interaction measure is
considerably larger. The second reason, perhaps more serious, lies in the
effect of the quark masses currently simulated. They are relatively heavy,
which increases the contribution of the quark condensate term to the
interaction measure, $\Delta_{m}(T)$ in the equation \eqref{eq:fullcondbag} 
below.

In the presence of massive quarks the trace of the energy-momentum tensor
takes the form from the trace anomaly \cite{CDJo} as follows:    
\begin{equation}
\langle \Theta^{\mu}_{\mu} \rangle ~=~ m_q\langle \bar{\psi}_q{\psi}_q \rangle~
       +~\langle G^2 \rangle,
\label{eq:thetaquark} 
\end{equation}
where $m_q$ is the light (renormalized) quark mass and ${\psi}_q$,
$\bar{\psi}_q$ represent the quark and antiquark fields respectively.  For the
sake of simplicity we choose two light quarks of the same mass $m_q = 6MeV$ and
respecting isospin symmetry.  We are now able to write down an
equation for the temperature dependence of the gluon condensate including the
effects of the light quarks in the trace anomaly in the following form:
\begin{equation}
\langle G^2 \rangle_T ~=~ \langle G^2 \rangle_0
~+~m_q\langle\bar{\psi}_q{\psi}_q \rangle_0~
~-~m_q\langle\bar{\psi}_q{\psi}_q\rangle_T~
~-~\langle \Theta^{\mu}_{\mu}\rangle_T.
\label{eq:fullcond}
\end{equation}
It is possible to see that at very low temperatures the
additional contribution to the temperature dependence from the quarks      
is rather insignificant. However, in the range where the chiral symmetry
is being restored there is an additional effect from the term
$\langle \bar{\psi}_q{\psi}_q\rangle_T$. Well above \tc\/ after the
chiral symmetry has been completely restored the only remaining
effect of the quark condensate is that of the vacuum. This term would 
contribute negatively to the gluon condensate. Thus we expect~\cite{BoMi} 
that for the light quarks the temperature dependence can be quite 
important around \tc\/. In fact we may rewrite the above equation 
using the known vacuum estimates for $\langle G^2 \rangle_0$ and
$m_q\langle \bar{\psi}\psi \rangle_0$  in the simpler form
\begin{equation}\langle G^2\rangle_T~=
~B_4~-~m_q\langle \bar{\psi}_q{\psi}_q \rangle_T~-~\Delta_{m}(T)T^4,
\label{eq:fullcondbag}
\end{equation}
where the constant $B_4$ has the value $0.01183\text{GeV}^4$
for the u and d quarks alone \cite{SVZ1}. The use of this value
should not imply the given accuracy-- in reality the accuracy is
still the known value of $0.012\text{GeV}^4$. However, it should imply that
it could be possible directly in the neighborhood of \tc\/ that the
temperature dependence from the quarks is sufficiently strong so that the
rise in this contribution from temperature compensates to a certain extent
for the strongly negative tendency of the pure gluon condensate~\cite{BoMi}.
For the case of pure $SU(N)$ we know that the values just below \tc\/
of \DT\/ are very small--that is, about the same size as
$m_q\langle \bar{\psi}_q{\psi}_q \rangle_0$. Thus, to the extent that
the chiral symmetry has not been completely restored, its effect on 
$\langle G^{2} \rangle_T$ will be present below \tc\/.
~\\
\indent
As an end to this sketch of the gluon condensate in QCD we will mention a few 
obvious points. Where in simulations on pure $SU(N_{c})$
gauge theory we could depend on considerable precision in the determination
of \tc\/ and $\Delta(T)$ as well as numerous other thermodynamical functions,
this is not the case for the theory with dynamical quarks. The statistics for 
the measurements are generally smaller. The determination of the temperature
scale is thereby hindered so that it is harder to clearly specify a given
quantity in terms of $T$. Thus, in general, we may state that
the accuracy for the full QCD is way down when compared to the computations  
of the pure lattice gauge theories. However, there is a point that arises
from the effect that the temperatures in full QCD are generally lower, so that
\DT\/ is much smaller~\cite{BoMi}. Here we can only speculate with the present
computations \cite{BKT94,MILCeos6,edwinqmfklat96}. Nevertheless, there could be
an indication of how the stability of the full QCD
keeps $\langle G^2 \rangle _T$ positive for $T~>~\tc$.
The condensates in full QCD have also been considered by Koch and
Brown~\cite{kochbrown}. However, the lattice measurements they used were not
obtained using a non-perturbative method, nor was the temperature scale
obtained from the full non-perturbative beta-function. 



\section{Dilatation Current}

We have noticed in the previous sections that the fact that the trace
of the energy momentum tensor does not vanish for the strong interactions
has important implications for the equation of state. Here we shall discuss
briefly some more theoretical results relating to $T^{\mu}_{\mu}$. This
arises with respect to the dilatation current as well as the special
conformal currents, which are not conserved. The dilatation current $D^{\mu}$
may be defined in terms of the position four-vector $x^{\mu}$ and the
energy momentum tensor $T^{\mu \nu}$ as simply $x_{\alpha}T^{\mu \alpha}$. 
In the case of general energy momentum conservation one can find~\cite{Jack}

\begin{equation}
\partial_{\mu} D^{\mu}~=~T^{\mu}_{\mu}.
\label{eq:dilcurr}
\end{equation} 

We now look into a volume in four dimensional space-time ${\cal V}_{4}$
containing all the quarks and gluons at a fixed temperature $T$ in equilibrium
~\cite{Mill2}.
The above equation \eqref{eq:dilcurr} holds when the energy momentum and all
the (color) currents are conserved over the surface $\partial {\cal V}_{4}$
of the properly oriented four-volume ${\cal V}_{4}$.
Then from \eqref{eq:dilcurr}
\begin{equation}
\oint_{\partial {\cal V}_{4}}{\cal D}_{\mu}dS^{\mu}~=
~\int_{{\cal V}_{4}} T^{\mu}_{\mu}dV_{4},
\label{eq:dyxleform}
\end{equation} 
where the {\it dyxle} three-form is  ${\cal D}_{\mu}dS^{\mu}$ on the
three dimensional surface $\partial {\cal V}_{4}$
dual to $D_{\mu}dx^{\mu}$ the dilatation current one-form~\cite{Flan}.
On the right hand side of \eqref{eq:dyxleform} the integrated form
~$\int_{{\cal V}_{4}} T^{\mu}_{\mu} dV_{4}$
is an action integral involving the equation of state. Since $T^{\mu}_{\mu}>0$,
the action integral is not zero. This action integral gets quantized with the
fields.

\section{Summary and Conclusions}

Absolutely central to our development of this discussion is the nontrivial
equation of state~\eqref{eq:state} with $T^{\mu}_{\mu}(T)~>~0$ for all
reachable temperatures. Thus a strongly interacting gas of
quarks and gluons in equilibrium is {\it never} an ideal gas.
~\\
\indent
The lattice simulations for pure $SU(N_c)$ provide not only an
accurate computation of the interaction measure $\Delta(T)$, but also
of the temperature scale due to the calculation of the 
beta function at finite temperature. These properties are all needed in the
computation of $\theta^{\mu}_{\mu}(T)$ in the previous sections. 
It is clear that the gluon condensate becomes negative in pure gauge
theory, and keeps dropping with increasing temperature. 
We have also given a discussion of these quantities in relation to full QCD.
In all cases considered above we may conclude  $T^{\mu}_{\mu}(T)$ always 
remains positive so that the divergence of both the dilitation and the 
conformal currents remain positive, and are thereby not conserved. 
Therefore, the scale and conformal symmetries remain broken at all 
temperatures for nonabelian lattice gauge theories after quantization.
~\\
\indent
    An important point for the physical meaning of these results 
is the comparison of the form $\DT$ with the corresponding term in
Bjorken's equation of state $p\Delta_{1}$~\cite{BjM}.
The definition of $\Delta_{1}$ relates to the change of the effective
number of degrees of freedom for which a first order phase transition
may appear. We also notice the general similarity in shape of $\Delta_{1}$
to the lattice interaction measure
with the main difference being that the peak in the lattice simulations 
~\cite{Eng4,Boyd} is slightly above $\tc$. Further comparison with
Bjorken's model could lead to a better physical understanding 
of the equation of state gotten from lattice simulations.



\section{Acknowledgements}

\medskip
The author would like to thank Rudolf Baier and Krzysztof Redlich for many 
very helpful discussions. Also a special thanks goes to Graham Boyd with
whom many of the computations were carried out. He is grateful to the 
Bielefeld group for providing their data, and especially to J\"urgen Engels
for the use of his programs and many valuable explanations of the lattice
results. For financial support he gratefully recognizes the Fulbright
Scholarship in Wroclaw and Helmut Satz in Bielefeld. The Pennsylvania
State University provided the sabbatical leave of absence. 



\end{document}